\documentclass[sigconf]{acmart}

\def\BibTeX{{\rm B\kern-.05em{\sc i\kern-.025em b}\kern-.08emT\kern-.1667em\lower.7ex\hbox{E}\kern-.125emX}}

\copyrightyear{2019}
\acmYear{2019}
\acmConference[ACM/IEEE]{Joint Conference on Digital Libraries}{June 2019}{Urbana-Champaign, Illinois USA}
\acmBooktitle{JCDL '19: Joint Conference on Digital Libraries, June 02--06, 2019, Urbana-Champaign, Illinois}
\acmPrice{}
\acmDOI{}
\acmISBN{}

\begin{document}

%
% The "title" command has an optional parameter, allowing the author to define a "short title" to be used in page headers.
\title[Shared Feelings]{Shared Feelings: Understanding Facebook Reactions to Scholarly Articles}

%
% The "author" command and its associated commands are used to define the authors and their affiliations.
% Of note is the shared affiliation of the first two authors, and the "authornote" and "authornotemark" commands
% used to denote shared contribution to the research.
\author{Cole Freeman}
%\authornote{Cole's author note}
\affiliation{%
  \institution{Northern Illinois University}
  %\streetaddress{1 Th{\o}rv{\"a}ld Circle}
  \city{DeKalb}
  \country{Illinois}}
\email{cole.freeman9@gmail.com}

\author{Mrinal Kanti Roy}
%\authornote{Equal Contribution}
\affiliation{%
  \institution{Northern Illinois University}
  %\streetaddress{P.O. Box 1212}
  \city{DeKalb}
  \state{Illinois}
  \postcode{43017-6221}
}
\email{mkantiroy@niu.edu}

\author{Michele Fattoruso}
%\authornote{The secretary disavows any knowledge of this author's actions.}
\affiliation{%
  \institution{Northern Illinois University}
  %\streetaddress{P.O. Box 1212}
  \city{DeKalb}
  \state{Illinois}
  \postcode{43017-6221}
}
\email{z1840898@students.niu.edu}

\author{Hamed Alhoori}
%\authornote{Cole's author note}
\affiliation{%
  \institution{Northern Illinois University}
  %\streetaddress{1 Th{\o}rv{\"a}ld Circle}
  \city{DeKalb}
  \country{Illinois}}
\email{alhoori@niu.edu}

%
% By default, the full list of authors will be used in the page headers. Often, this list is too long, and will overlap
% other information printed in the page headers. This command allows the author to define a more concise list
% of authors' names for this purpose.
\renewcommand{\shortauthors}{Freeman et al.}

%
% The abstract is a short summary of the work to be presented in the article.
\begin{abstract}
Research on social-media platforms has tended to rely on textual analysis to perform research tasks. While text-based approaches have significantly increased our understanding of online behavior and social dynamics, they overlook features on these platforms that have grown in prominence in the past few years: click-based responses to content. In this paper, we present a new dataset of Facebook Reactions to scholarly content. We give an overview of its structure, analyze some of the statistical trends in the data, and use it to train and test two supervised learning algorithms. Our preliminary tests suggest the presence of stratification in the number of users following pages, divisions that seem to fall in line with distinctions in the subject matter of those pages.
\end{abstract}

%\begin{CCSXML}
%<ccs2012>
%<concept>
%<concept_id>10002951.10003227.10003351</concept_id>
%<concept_desc>Information systems~Data mining</concept_desc>
%<concept_significance>500</concept_significance>
%</concept>
%<concept>
%<concept_id>10002951.10003260.10003282.10003292</concept_id>
%<concept_desc>Information systems~Social networks</concept_desc>
%<concept_significance>500</concept_significance>
%</concept>
%<concept>
%<concept_id>10002951.10003227.10003241.10003244</concept_id>
%<concept_desc>Information systems~Data analytics</concept_desc>
%<concept_significance>300</concept_significance>
%</concept>
%</ccs2012>
%\end{CCSXML}

%\ccsdesc[500]{Information systems~Data mining}
%\ccsdesc[500]{Information systems~Social networks}
%\ccsdesc[300]{Information systems~Data analytics}

%
% Keywords. The author(s) should pick words that accurately describe the work being
% presented. Separate the keywords with commas.
\keywords{Facebook Reactions, Altmetrics, Data Collection, Social Clicks, Research Community, Social Media Analytics, Supervised Learning}

%
% A "teaser" image appears between the author and affiliation information and the body
% of the document, and typically spans the page.
%%  \includegraphics[width=\textwidth]{sampleteaser}
%%  \caption{Seattle Mariners at Spring Training, 2010.}
%%  \Description{Enjoying the baseball game from the third-base seats. Ichiro Suzuki preparing to bat.}

%
% This command processes the author and affiliation and title information and builds
% the first part of the formatted document.
\maketitle

\section{Introduction}

% Introduce problem
% What can Facebook Reactions tell us about scholarly articles?

%An increasing amount of scholarly content is being shared and discussed daily on online platforms. Whereas traditional citations measure impact within scholarly boundaries, alternative metrics or altmetrics [1] can be used to measure a broader range of influences.

As the prevalence of social media in the world around us increases and the number of users on these online platforms grows, so too grows the rate at which scholarly content is being proliferated and discussed in these venues. More and more, academics are finding it rewarding to look to these platforms for the insight they provide into research problems. 

One reason scholars have turned to social media is to measure the influence their work is having in those spaces; this has become known as alternative metrics, or \textit{altmetrics}~\cite{Alhoori:2014:AFC:2740769.2740833}; another reason is for the knowledge online platforms provide about human behavior--an area of research known as \textit{social-media analytics}~\cite{krebs:social_emotion_mining_reaction_prediction, bayer_ellison_measuring_emotion, Rajadesingan:2015:SDT:2684822.2685316}. Studies in social-media analytics tend to focus either on text, using approaches such as Natural Language Processing (NLP), sentiment analysis, or opinion mining to arrive at and support research conclusions~\cite{W17-1102}, or on the proliferation of content through online communities~\cite{gabielkov:hal-01281190}. These approaches have proved effective for understanding or predicting many aspects of human behavior; but they leave a number of other expressive signals unexamined. 

Click-based reactions, on the other hand, are a relatively underutilized resource in social-media research. Examples of quick-draw, ready-made expressive features are becoming increasingly prevalent across many platforms, and as such have attracted some amount of attention from researchers in the past few years~\cite{pool:emotion_detection, basile2017}.

In this paper, we present a new dataset of click-based reactions to scholarly articles on Facebook and use it to gain insight into how users are interacting with scholarly articles on that platform. In addition to information about the articles themselves, our dataset records the count of each click-based feature we could access through Facebook's Graph API. We use our newly developed dataset to train and test two machine learning algorithms, and our analysis of the results shines some light on surprising relationships between features.

\section{Building the dataset}

Before going any further, it will be useful to define a few terms and features:

\begin{itemize}
    \item \textbf{Click-based reactions -} non-textual user interactions with shared content--sometimes referred to simply as \textit{reactions}; includes Facebook Likes and Reactions, Re-shares, and Page visibility (definitions for these last two are below).
    \item \textbf{Reactions -} the five click-based reactions: Love, Amazed, Laughing, Sad, and Angry; will be distinguished from the common term ``reaction'' by capitalization.
    \item \textbf{Page visibility -} the number of followers a Facebook page has.
    \item \textbf{Re-shares -} the number of times users have re-shared a public post of an article into another location.
\end{itemize}

% ALTMETRIC
The roots of our dataset lie in the online resource Altmetric\footnote{\url{https://www.altmetric.com}.}, which tracks the impact scholarly articles have across a variety of social media platforms. We used Altmetric as a ``jumping-off point'', querying their API\footnote{\url{http://api.altmetric.com/}.} for information on articles we were interested in and for the public pages onto which they had been shared. It gave us access to the titles, publication dates, subjects, and the URLs of Facebook shares for nearly 1.5 million scholarly articles.

We targeted content shared on Facebook rather than other social-media platforms for several reasons. First, Facebook offers its users a variety of click-based interactions with which they can personalize their response to content; other platforms we considered targeting such as Twitter have more limited palettes of click-based reactions available to users. Second, Facebook's enormous population of active users increases the likelihood that content shared there will receive more attention: it has about 2.27 billion active monthly users, almost seven times Twitter's active population of 330 million. Third, the impact of scholarly articles on Twitter users has been the subject of many recent studies~\cite{priem_cite_on_twitter, Shuai_preprints_twitter_mentions, gabielkov:hal-01281190}, whereas the response to this type of content on Facebook remains largely unexamined.

With our list of Facebook URLs for article shares, we queried Facebook's Graph API\footnote{\url{https://developers.facebook.com/docs/graph-api/}.} for the reaction counts on each post. Our dataset records their responses, and was collected during the period of December 1-13, 2018. Constraints in the number of queries allowed by Facebook's API (200/hour) determined the rate at which we could work. The resulting dataset is publicly available on OSF\footnote{\url{https://osf.io/4kh7r/}.} as a comma-separated-value file (CSV).

% SCOPE OF DATASET
We limited our collection efforts only toward scholarly articles published in 2017. Choosing this year accomplished three goals: (i) Reactions were released by Facebook in February 2016~\cite{fb-reactions:release}, so any articles we looked at had to be published after that time to have meaningful data on this feature; (ii) any time a new feature is unrolled, it takes some amount of time for users to learn how to use it; Prah~\cite{fb-reactions_use} finds that use of Reactions increased from 2.4\% of all interactions in April 2016 to 5.8\% by June 2016, and up to 12.8\% of all interactions by June 2018; by the time of our data collection in December 2018 a large enough subset of users were comfortable expressing themselves with the feature to warrant more scholarly attention; and (iii) by the time we began our data collection, a sufficient interval of time had passed for articles to be widely shared and reacted to (between 11 and 23 months). 

\begin{figure}
  \includegraphics[width=0.45\textwidth]{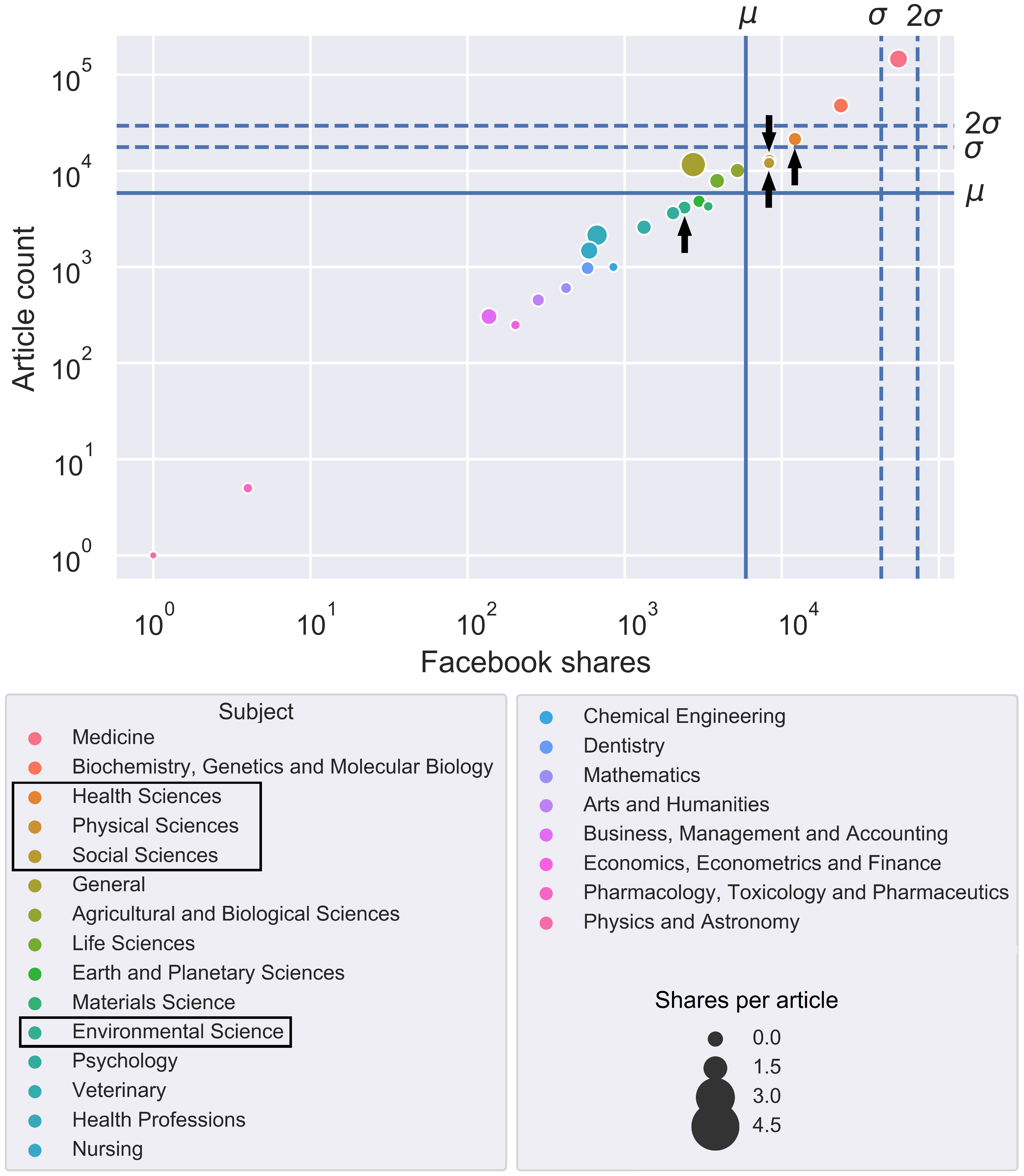}
  \caption{Count of articles published in 2017 categorized by subject plotted against number of shares. Both features are shown on logarithmic axes.}
  \label{fig:subjects_scatter}
\end{figure}

Of all the articles tracked by Altmetric, we found 296,052 were published in 2017 and had been shared on Facebook at least once. We eliminated entries that were missing data on the pages to which the articles had been shared; this reduced our set to 135,635 articles. We further limited the scope by selecting only articles with Scopus\footnote{\url{https://www.scopus.com/}.} subjects in the scientific domain. We chose to focus only on articles in the Health Sciences, Physical Sciences, Social Sciences, and Environmental Science. Figure~\ref{fig:subjects_scatter} shows that these four categories, article counts fall within one standard deviation of the mean number of articles, as do the total number of Facebook shares (Health sciences is the only exception, exceeding one standard deviation greater than the mean of article counts). It also displays the full list of subjects in all the 2017 articles and gives a sense of their distribution. The mean and two standard deviations are indicated there with blue lines for both axes, and the four subjects we target are indicated with arrows in the plot. Limiting the scope of subjects reduced the number of articles needed to process to just over 31,000. When we removed articles with missing features such as abstract and title, we were left with 11,474 articles: these are the articles recorded in our dataset.

% PRIVACY CONCERNS --- POLICY CONCERNS
In our data collection process, we took the utmost care to respect Altmetric's and Facebook's specifications for \textit{how} and \textit{why} their data can be accessed and used and to protect the personal information of social-media users. Our interests are only in the ways that people are interacting \textit{in the aggregate} with scholarly content on social media platforms--not in specific ways users' beliefs or opinions may influence their behavior. We recognize that identifying information could in some instances be inferred \textit{a posteriori} from some of the data we collect; however, our method of data collection does not target anything that could be used to consistently identify individual users and avoids collecting identifying information about individuals.

\section{Data Exploration}

\begin{figure}
  \includegraphics[width=0.45\textwidth]{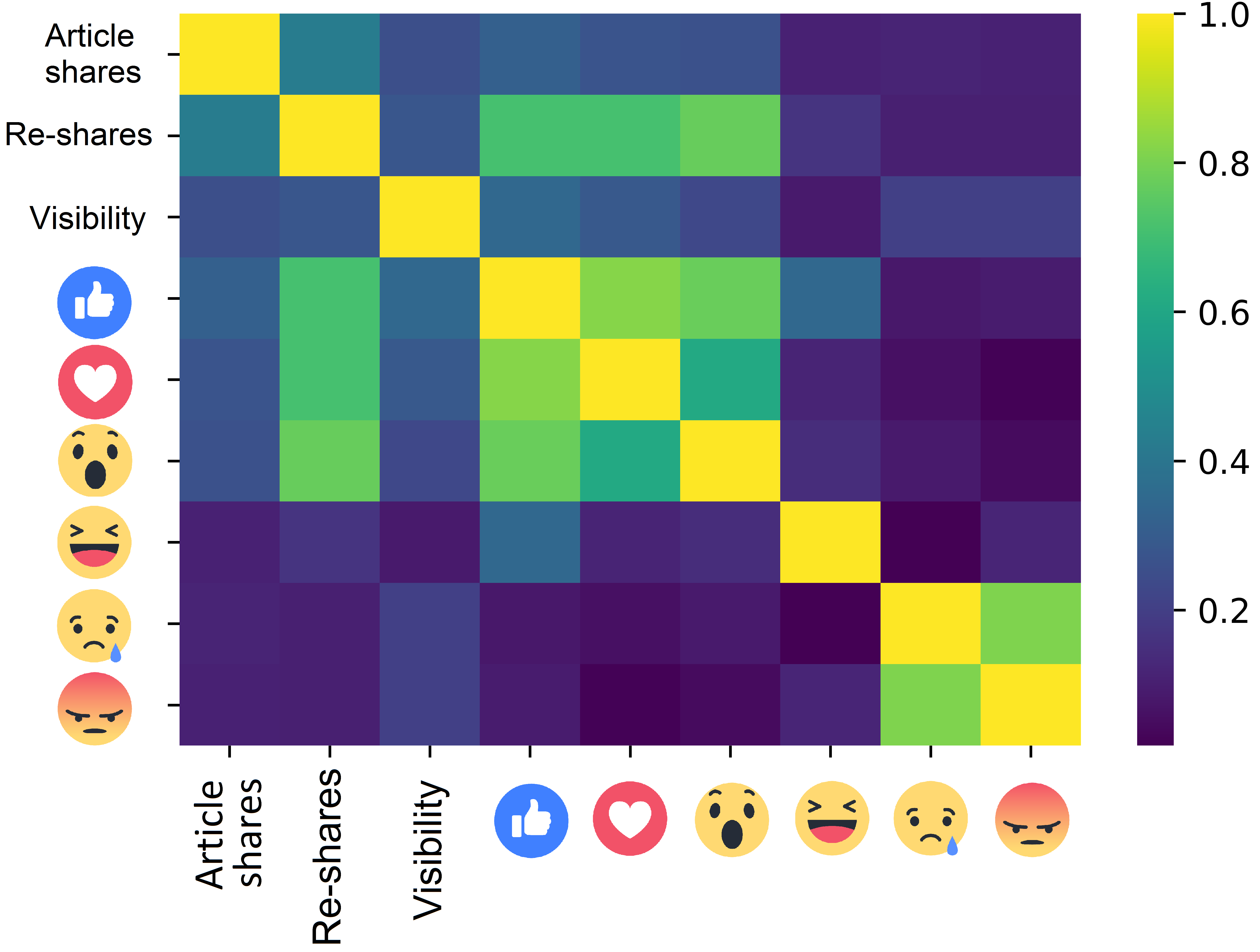}
  \caption{Correlation matrix of features using Pearson's $r$.}
  \label{fig:feat_corr_matrix}
\end{figure}

The click-based features of our dataset are displayed along the axes in Figure~\ref{fig:feat_corr_matrix}; also displayed are the Pearson $r$ correlation coefficients for all feature pairs. Highly correlated pairs are: Like and Love ($r=0.82$), Sad and Angry ($r=0.81$), Like and Amazed ($r=0.77$), Love and Re-shares ($r=0.71$). We can infer that high positive correlation is a sign that users employ features in similar contexts, and that the emotional expressions represented by those features overlap. For example, a Like seems to have a meaning comparable to a Love or (to a lesser extent) an Amazed, or (to an even lesser extent) a Laughing reaction. These relationships may not surprise us because they are all positive emotional states; but other feature pairs that have related expressive values in usage, such as Angry and Sad reactions, are not so intuitive.

Likes and Re-shares are correlated with the most other features; this might be explained by the fact that these two are the oldest reactions--but we also notice they are correlated with other \textit{emotionally positive} reactions such as Love or Amazed and not with the negative emotions Sad or Angry. It follows that by Liking or Re-sharing a post, a user expresses a positive emotional reaction to its content. Looking at this from another angle, we infer that content that is more likely to inspire a negative reaction from users is less likely to be Re-shared or Liked.

High correlation between features can lead to increased variance in model results. To counter this, modelers often eliminate one of a pair of correlated features. Rather than removing features and losing data in our sparse dataset, we combined Love/Wow and Sad/Angry Reactions into two new composite features for our models. 

Low correlation signifies that features have relatively distinct use values. Among the lowest $r$ coefficients are Love/Angry ($r=0.145$) and Laughing/Sad ($r=0.224$); this makes intuitive sense, as these reactions nominally encompass opposite emotions. Laughing/Page visibility ($r=0.229$) is another low-correlation pair, suggesting that articles that inspire humor are more likely to be posted to public pages with relatively low follower counts. It is likely that this relationship may be a result of our choice to limit the articles we include to those in the scientific domain, where humor is an under-utilized affect. 

\begin{figure}
  \includegraphics[width=0.45\textwidth]{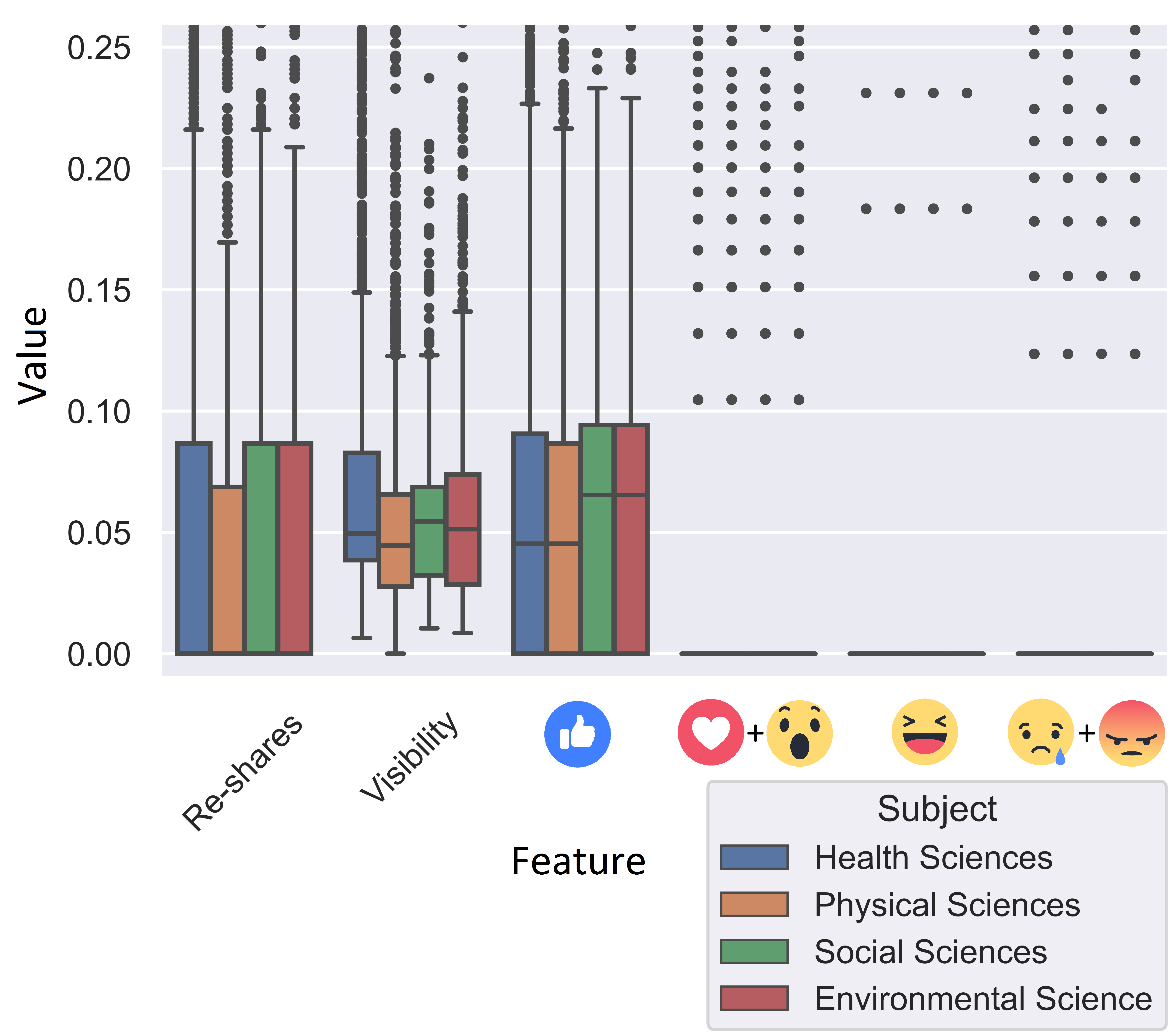}
  \caption{Feature values after transformation displayed by subject. Values only go up to 0.25 to better display the IQRs.}
  \label{fig:feat_boxplot}
\end{figure}

Our dataset also contains outliers in each feature category; to correct for these, we re-scaled the features to a range from 0 to 1, then took the cube root to these new values. Our root normalization function is demonstrated in Equation~\ref{eq:root_normalization}; it helped to smooth the distribution of values, raising the lower values by more than it raised the higher values. The result after combination/normalization is displayed in Figure~\ref{fig:feat_boxplot}. 

\begin{equation}
  rt\_norm(F_{i}) = \sqrt[\leftroot{-1}\uproot{2}\scriptstyle 3]{\frac{F_{i} - F_{min}}{F_{max} - F_{min}}}
  \label{eq:root_normalization}
\end{equation}

Even after transformation, our dataset is still sparse--zeros are un-changed by the transformation; yet features with greater variance, such as Visibility or Likes, have less spread between the IQR and outliers. The median value of all Reactions is zero, and non-zero values in those features are all in the fourth quartile. Likes have the largest interquartile range (IQR), though the median is still close to zero. Page visibility and Likes have the highest median values of all features.

\section{Supervised-learning models}

% Details about how we trained/tested our models
To explore the relationships in our dataset further, we isolated two feature subsets and trained two supervised learning classification algorithms with them. We used Decision Tree and Random Forest algorithms because of the insight they provide into the relationships between features, and our feature sets are detailed in Table~\ref{table:model_feature_sets}. We were interested in gaining insight into the extent to which users' interactions could be related to articles' subject matter; and so we selected article subjects as the class labels for our models. This gave us four targets for our multiclass classification models to predict. 

\begin{table}[]
\centering
\resizebox{\columnwidth}{!}{%
\begin{tabular}{|c|c|c|}
\hline
\textbf{Set} & \textbf{Features}                                                                             & \textbf{Model Accuracy/AUC}                                                                           \\ \hline
\textbf{A}   & \begin{tabular}[c]{@{}c@{}}Likes, Re-shares, Love/Amazed, \\ Laughing, Sad/Angry\end{tabular} & \begin{tabular}[c]{@{}c@{}}Random Forest: 38.1\% / 67\%\\ Decision Tree: 40.97\% / 65\%\end{tabular}  \\ \hline
\textbf{B}   & \begin{tabular}[c]{@{}c@{}}Visibility, Love/Amazed, \\ Laughing, Sad/Angry\end{tabular}       & \begin{tabular}[c]{@{}c@{}}Random Forest: 64.21\% / 82\%\\ Decision Tree: 65.66\% / 77\%\end{tabular} \\ \hline
\end{tabular}
}
\caption{The two feature sets used in our classification models, with model accuracies and Areas Under the Curve (AUC).}
\label{table:model_feature_sets}
\end{table}

With the first set (\textbf{A}) we were interested in finding the extent to which click-based reactions that are immediately available to users on the post itself could be used to estimate an article's subject. The second set (\textbf{B}) provides insight into how extended click-based features such as Page Visibility can be used to approximate the subject matter of posts. 

\begin{figure}
  \includegraphics[width=0.45\textwidth]{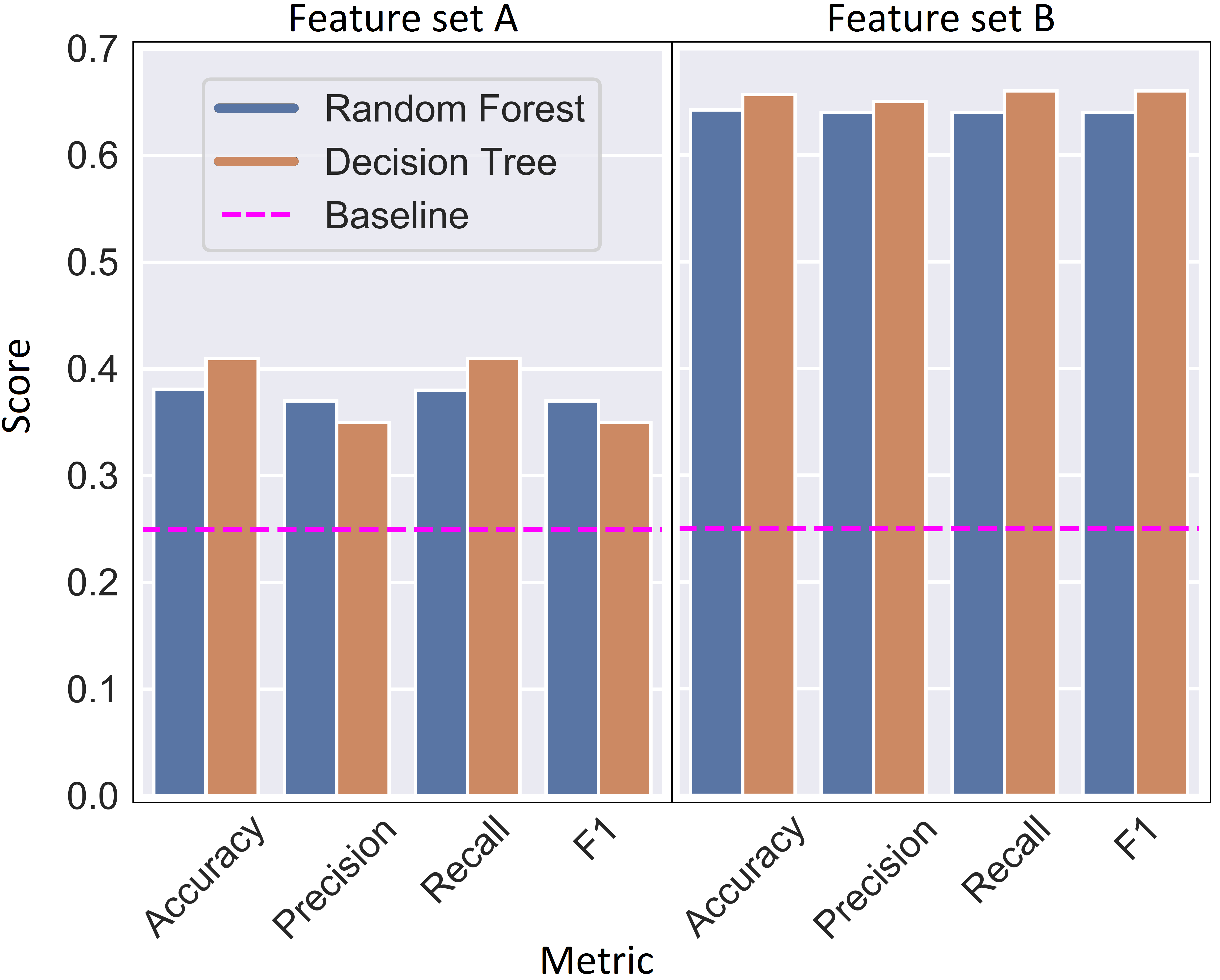}
  \caption{Various metrics used to score our two ML algorithms applied to the two feature sets. Baseline represents the results of random guesses at article subjects.}
  \label{fig:model_scores}
\end{figure}

Table~\ref{table:model_feature_sets} displays the accuracy and Area Under the Curve (AUC) of our models, and Figure~\ref{fig:model_scores} shows the results of our models using several different metrics. For reference, scores are shown against the baseline, which represents random guesses at which of the four class labels an article belongs to. Feature set B produced significantly better results than A with both algorithms. Average accuracy of models with feature set B is 160\% greater than the baseline, while feature set A is only 58\% greater.

\begin{figure}
  \includegraphics[width=0.45\textwidth]{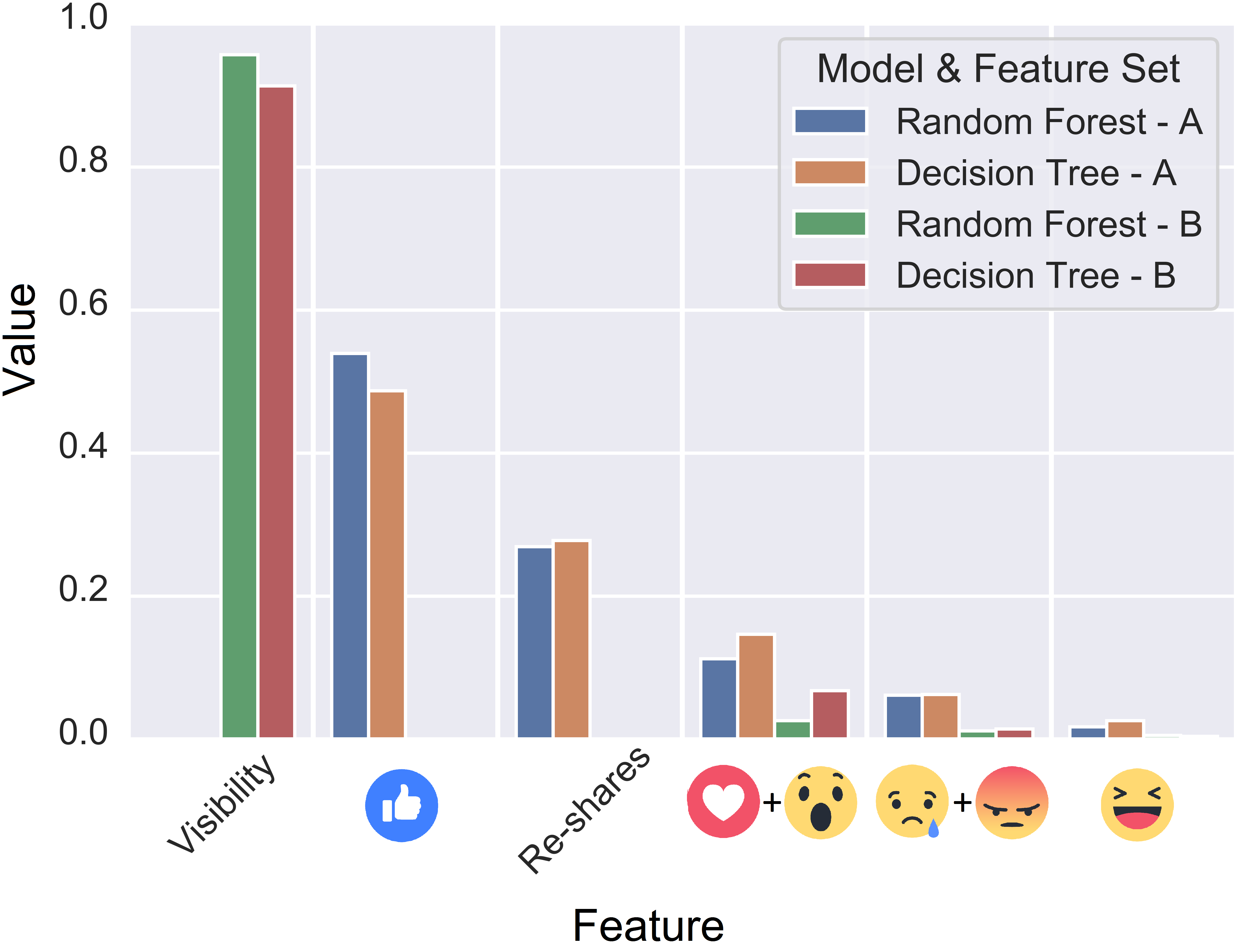}
  \caption{Feature weight for the four implemented models (two algorithms $\times$ two feature sets).}
  \label{fig:feat_importance}
\end{figure}

Figure~\ref{fig:feat_importance} shows the relative importance of each feature in our models. In feature set A, Likes have the greatest weight, accounting for 51\% of the result on average between the two algorithms; the weight of Re-shares is the second highest importance, accounting for an average of 27\% of the result. In feature set B, Visibility is the most important feature accounting for an average of 94\% of the result; the remainder of the weight is spread relatively evenly among the remaining features. 

\section{Discussion and Conclusion}

%Our analysis shines some light on possible connections between features in our dataset of click-based reactions on Facebook.

%Our purpose in implementing these models was to shine some further light on the relationships between the features, as well as to direct future research endeavors using our new dataset.

%A take-away from our experiments with ML algorithms might be not so much that click-based reactions are predictive of article subject matter, but that the visibility of pages by subject matter is highly tiered.

% FUTURE WORK

%It is worth noting that assumption we have made that is worth testing in future work is that users can determine the subject matter of a post by the title and text provided by the sharer.

%future work will involve both increasing its size and scope, as well as finding collection methods that introduce less bias in this regard

Our new dataset of click-based reactions to scholarly content on Facebook offers a wealth of possibilities for researchers interested in social media analytics. We have demonstrated how it can be used in the exploration of user interactions with scholarly content on Facebook, and how click-based reactions are an effective data source for investigating indicators of user emotional attitudes.

Results from the models trained and tested on our dataset suggest that the number of followers a page has (Visibility) may be predictive of article subject matter; this indicates that there may be patterns in the content shared on Facebook pages and the number of followers these pages have. It may prove useful for researchers to explore the ways in which Facebook page popularity is stratified by the type of content displayed on their pages. 

We have also suggested some interpretation of Facebook click-based reactions that are not immediately apparent, notably that Re-shares convey an emotionally positive feelings toward content, and that Sad and Angry Reactions express similar affects. These relationships are not at all obvious, and give us insight into how these features are being used in practice. 

%\begin{acks}
%  The authors would like to thank Dr. Hamed Alhoori for his time and attention given to different matters of this project. 
  
%  The authors would also like to thank Altmetrics for their voluminous data on scholarly articles, without which collecting our data would have been a near insurmountable task. We would also like to thank Facebook for their access to their Graph API, which provided us with the ability to access data about users reactions to posts on public pages. We also thank the authors of scikit-learn, Pedregosa et al. \cite{scikit-learn}.

%\end{acks}

%
% The next two lines define the bibliography style to be used, and the bibliography file.
\bibliographystyle{ACM-Reference-Format}
\bibliography{sample-base}

%
% If your work has an appendix, this is the place to put it.

\end{document}